# A fast-converging scheme for the Phonon Boltzmann equation with dual relaxation times

Jia Liu, Lei Wu*

Department of Mechanics and Aerospace Engineering, Southern University of Science and Technology, Shenzhen 518055, China

* Corresponding author, E-mail: wul@sustech.edu.cn

**Abstract**

Callaway's dual relaxation times model, which takes into account the normal and resistive scatterings of phonon, is used to describe the heat conduction in materials like graphene. For steady-state problems, the Callaway model is usually solved by the conventional iterative scheme (CIS), which is efficient in the ballistic regime, but inefficient in the diffusive/hydrodynamic regime. In this paper, a general synthetic iterative scheme (GSIS) is proposed to expedite the convergence to steady-state solutions. First, macroscopic synthetic equations are designed to guide the evolution of equilibrium distribution functions for normal and resistive scatterings, so that fast convergence can be achieved even in the diffusive/hydrodynamic regime. Second, the Fourier stability analysis is conducted to find the convergence rate for both CIS and GSIS, which rigorously proves the efficiency of GSIS over CIS. Finally, several numerical simulations are carried out to demonstrate the accuracy and efficiency of GSIS, where up to three orders of magnitude of convergence acceleration is achieved.

**Keywords** : Callaway model, Discrete velocity method, Fourier stability analysis, general synthetic iterative scheme, Phonon Boltzmann equation

## 1. Introduction

With the rapid development of semiconductor technology, the characteristic size of microelectronic devices is getting smaller and smaller. And the heat conduction in micro- and nano-scale devices has attracted extensive attention[1], as it has strong applications in many fields, including the device heat dissipation control and thermal functional material development. Although the classical Fourier's heat conduction law has been guiding the study of heat conduction in past centuries, it is no longer valid at the microscale and nanoscale. When the Knudsen number is not small (i.e., the characteristic length is comparable to or even smaller than the mean free path of the phonon[2]), the Boltzmann transport equation (BTE), rather than Fourier's law, is adopted to describe the multiscale heat conduction, from the diffusive/hydrodynamic regime to the ballistic regime.

The phonon BTE is a nonlinear integrodifferential equation defined in a high-dimensional phase-space (i.e., time $t$, physical space $\boldsymbol{x}$, solid angle $\Omega$, and frequency $\omega$), so its numerical simulation poses a great research challenge. Similar to gas kinetic models, the relaxation time approximation is introduced to simplify the scattering term in phonon BTE: the smaller the relaxation time, the stronger the scattering. For example, the single-mode relaxation time approximation[3][4] is commonly used to describe the resistive scattering (conserves energy but not momentum), which has been shown to well predict the heat conduction in silicon and germanium. But for materials in which the normal scattering (conserves energy and momentum) plays a significant role, say, graphene, this approximation is not applicable. To circumvent this problem, Callaway's dual relaxation model[5] is proposed, which assumes that the normal and resistive scatterings restore the phonon distribution to a displaced Planck distribution and a Planck distribution, respectively.

There are two ways to solve the Callaway model. The first is the derivation of macroscopic equations, while the second is the direction numerical simulation. For the former, Guyer and Krumhansl[6] first obtained the GK equations from the BTE. Subsequently, macroscopic equations have been derived via the Chapman-Enskog expansion[7] and moment method[8]. However, these macroscopic equations are only accurate near the diffusive/hydrodynamic regime. For example, Alvarez, Jou, and Sellitto used the classic phonon hydrodynamic equations with slip boundary condition to explain the heat conduction in the micro-nano scale[9]. For the later, efficient and accurate numerical methods are urgently needed since the BTE is defined in high-dimensional phase-space.

Many numerical methods have been developed to solve the BTE, including the Monte Carlo scheme[10],

lattice Boltzmann method[11], discrete ordinate method[12][13], finite volume method[14], and discrete unified gas kinetic scheme[15][16]. The Monte Carlo method is a stochastic method which uses simulated particles to represent the advection and scattering of phonons. Since the advection and scattering are splitted, it is required that the time step should be smaller than the relaxation time, and the grid size smaller than the phonon mean free path. As a result, the Monte Carlo method is expensive in the diffusive/hydrodynamic regime[17]. Moreover, it suffers from large statistical errors due to its stochastic nature, especially when the temperature variation is small. To mitigate the latter problem, the variance-reduced Monte Carlo method was proposed, in which only the deviation from equilibrium are simulated[18]. The standard lattice Boltzmann method, initially proposed for fluid dynamics, is extended to solve the phonon BTE numerically. Since it is based on the near-equilibrium hypothesis, it is difficult to handle strong nonequilibrium effects and phonon spectral properties. Hence, it is valid for small Knudsen numbers but is incapable of capturing nonequilibrium heat conduction. The discrete ordinate method and finite volume method are all based on the discrete-ordinate formulation, where the solid angle is discretized. However, the numerical dissipation in the diffusive/hydrodynamic regime is large, since, like the Monte Carlo method, the advection and scattering are handled separately. This problem is fixed in the discrete unified gas kinetic scheme, where the advection and scattering are handled simultaneously, so that large spatial cell size and time step can be used.

Although great progress has been made in the past decades, the explicit marching methods are still inefficient when dealing with steady heat conduction problems. For example, when the normal scattering is strong while the resistive scattering is weak, the temperature profile changes rapidly in the vicinity of solid walls, while in the bulk region the temperature variation is small. In this case, non-uniform spatial grids should be used to reduce the computational cost. However, due to the Courant–Friedrichs–Lewy (CFL) condition, the time step is usually limited, and it usually takes a huge number of time steps to find the steady-state solutions[19]. Therefore, the implicit numerical method is highly demanded. To this end, the conventional iterative scheme (CIS) is a simple and common implicit method to obtain the steady-state solution of phonon BTE, where the scattering operator is split into the gain term and the loss term, and the time derivative is dropped. The advection operator and loss term are calculated at the current iteration step, while the gain term is evaluated at the previous iteration step. When the Knudsen number is large, the CIS is efficient since the converged solution can be obtained within a few iterations. However, the iteration number increases sharply when Knudsen number is small; worse still, the spatial grid size should be smaller than the mean free path to keep the numerical dissipation low.

Since the above-mentioned method cannot quickly obtain numerical results for heat conduction with a wide range of Knudsen numbers, it is urgent to develop some acceleration methods to improve the convergence efficiency in the diffusive regime and ballistic regime. If the normal scattering is absent, the phonon BTE has the same form as that for the radiation transport, where in the diffusive regime many acceleration methods have been developed to find the steady-state solution, e.g., the synthetic iterative scheme[20]. The basic idea of synthetic iterative scheme is that the BTE and macroscopic equations (diffusion equation when only resistive scattering is considered) are solved simultaneously: the mesoscopic BTE provides high-order moments to macroscopic equations, while the macroscopic equations provide macroscopic quantities appearing in the scattering term. Since the diffusion equation allows very efficient exchange of information, fast convergence is achieved, which can find the converged solution within dozens of iterations.

However, for the Callaway's model, the limiting equation is not the diffusion equation, but is described by the GK equation[6]. Therefore, the design and analysis of the synthetic iteration scheme will be different to that of resistive scattering only. It is therefore the main aim in the present work to develop an efficient numerical scheme for the Callaway model.

The remainder of this paper is organized as follows: in section 2, the CIS and general synthetic iteration scheme (GSIS) for the phonon Boltzmann equation under Callaway's model are introduced. Specifically, the convergence rates of CIS and GSIS are calculated rigorously by the Fourier stability analysis; in section 3, several numerical simulations in different phonon transport regimes are conducted to demonstrate the fast-convergence property of GSIS, and the effective thermal conductivity is analysed; the summary and outlooks are given in section 4.

## 2. Kinetic model and numerical methods

### 2.1 The Callaway model

The phonon BTE can be used to describe the heat conduction in solid materials, but the scattering term is extremely complicated. In general, the phonon BTE is simplified with the dual relaxation time approximation, resulting in the following Callaway model:

$$\frac{\partial f}{\partial t} + \boldsymbol{v} \cdot \boldsymbol{\nabla}_x f = \frac{f_{eq}^R - f}{\tau_R} + \frac{f_{eq}^N - f}{\tau_N}, \tag{1}$$

where $f$ is the phonon distribution function that depends on the time $t$, wave vector $\boldsymbol{k}$, and spatial coordinates $\boldsymbol{x}$. The phonon group velocity is

$$\boldsymbol{v} = \boldsymbol{\nabla}_\mathbf{k} \omega = |\boldsymbol{v}|\boldsymbol{s}, \tag{2}$$

where $\omega$ is the frequency, and $\boldsymbol{s} = (\cos\theta, \sin\theta\cos\varphi, \sin\theta\sin\varphi)$ is the direction vector. For a phonon with given wave vector, the frequency $\omega$ is determined from the dispersion relation $\omega(\boldsymbol{k})$. For the following discussion, it is assumed that $|\boldsymbol{v}| = 1$. $\tau_R$ and $\tau_N$ denote the mean free time of the phonon resistive and normal scatterings, respectively. Here, the gray-matter assumption is adopted so that the relaxation times $\tau_R$ and $\tau_N$ are independent of the wave vector or frequency. $f_{eq}^R$ is the equilibrium distribution function in the resistive process, following the Bose-Einstein distribution[21],

$$f_{eq}^R = \frac{1}{\exp\left(\frac{\hbar\omega}{k_B T}\right) - 1}, \tag{3}$$

while $f_{eq}^N$ is the equilibrium distribution function in the normal process, following the drifted Bose-Einstein distribution,

$$f_{eq}^N = \frac{1}{\exp\left(\frac{\hbar\omega - \hbar\boldsymbol{k}\cdot\boldsymbol{u}}{k_B T}\right) - 1}, \tag{4}$$

where $\hbar$ is the Planck's constant divided by $2\pi$, $k_B$ is the Boltzmann constant, $T$ and $\boldsymbol{u}$ (related to heat flux) are the macroscopic temperature and drift velocity, respectively.

Equation (1) can also be expressed in terms of the phonon energy density per unit solid angle $e(\boldsymbol{x}, \boldsymbol{s}, t)$[17],

$$\frac{\partial e}{\partial t} + \boldsymbol{v} \cdot \boldsymbol{\nabla} e = \frac{e_{eq}^R - e}{\tau_R} + \frac{e_{eq}^N - e}{\tau_N}, \tag{5}$$

where

$$e(\boldsymbol{x}, \boldsymbol{s}, t) = \sum_p \int \hbar\omega f(\omega) D_p(\omega) d\omega, \tag{6}$$

with $D_p(\omega)$ being the phonon density of state, and the subscript $p$ representing the phonon polarization. When the temperature variation is small, the equilibrium distribution functions of the resistive and normal scatterings can be linearized, respectively, as

$$\begin{aligned} e_{eq}^R &= \frac{C_v T}{4\pi}, \\ e_{eq}^N &= \frac{C_v T}{4\pi} + \frac{C_v T_0 \boldsymbol{u} \cdot \boldsymbol{s}}{4\pi|\boldsymbol{v}|}, \end{aligned} \tag{7}$$

where $C_v$ is the volumetric specific heat, $T$ is the deviation temperature from the reference temperature $T_0$. According to the facts that the energy is conserved in both resistive and normal scatterings, and the momentum is conserved in the normal scattering, i.e.,

$$\int \frac{e_{eq}^R - e}{\tau_R} d\Omega = 0,$$
$$\int \frac{e_{eq}^N - e}{\tau_N} d\Omega = 0, \qquad \int \frac{e_{eq}^N - e}{\tau_N} \boldsymbol{v} \, d\Omega = 0, \tag{8}$$

the corresponding macroscopic variables (the temperature $T$, drift velocity $\boldsymbol{u}$ and heat flux $\boldsymbol{q}$) can be calculated from $e$ as follows,

$$T = \frac{1}{C_v} \int e \, d\Omega, \quad \boldsymbol{u} = \frac{3}{C_v T_0} \int \boldsymbol{v} e \, d\Omega, \quad \boldsymbol{q} = \int \boldsymbol{v} e \, d\Omega. \tag{9}$$

### 2.2 Moment equations

We are interested in the steady-state solution of phonon Boltzmann equation, so the term related to time derivation is ignored. On multiplying Eq.(5) by 1, $v_i$, $v_i v_j$, respectively, and integrating the resultant equations with respect to the solid angle $\Omega$, we obtain the following trace-free moments equations for the evolution of temperature and heat flux[8]:

$$\frac{\partial q_k}{\partial x_k} = 0,$$
$$\frac{C_v}{3} \frac{\partial T}{\partial x_i} + \frac{\partial N_{\langle ik \rangle}}{\partial x_k} = -\frac{1}{\tau_R} q_i, \tag{10}$$
$$\frac{2}{5} \frac{\partial q_{\langle i}}{\partial x_{j \rangle}} + \frac{\partial M_{\langle ijk \rangle}}{\partial x_k} = -\frac{1}{\tau_C} N_{\langle ij \rangle}.$$

where

$$N_{\langle ik \rangle} = u_{ik} - \frac{1}{3} \delta_{ik} u_{hh} = \int v_i v_k e \, d\Omega - \frac{1}{3} \delta_{ik} \int v_h v_h e \, d\Omega = \int \left( v_i v_k - \frac{1}{3} \delta_{ik} \right) e \, d\Omega,$$
$$\frac{\partial q_{\langle i}}{\partial x_{j \rangle}} = \frac{1}{2} \frac{\partial q_i}{\partial x_j} + \frac{1}{2} \frac{\partial q_j}{\partial x_i} - \frac{1}{3} \frac{\partial q_k}{\partial x_k} \delta_{ij},$$
$$M_{\langle ijk \rangle} = u_{ijk} - \frac{1}{5} \left( u_{ihh} \delta_{jk} + u_{jhh} \delta_{ik} + u_{khh} \delta_{ij} \right) = \int v_i v_j v_k e \, d\Omega - \frac{1}{5} \left[ \int (v_i \delta_{jk} + v_j \delta_{ik} + v_k \delta_{ij}) e \, d\Omega \right].$$

The above moment equations are not closed because the higher-order moments are not known a priori. Inspired by the Grad's moment method, Fryer and Struchtrup[8] closed the moment equations by approximating the distribution function by the principle of maximum phonon entropy. In this case, the term $\frac{\partial M_{\langle ijk \rangle}}{\partial x_k}$ vanishes, and the above momentum equations are reduced to the Guyer-Krumhansl equations for phonon hydrodynamics:

$$\frac{\partial q_k}{\partial x_k} = 0,$$
$$\frac{C_v}{3} \frac{\partial T}{\partial x_i} - \frac{1}{5} \tau_C \left( \frac{\partial^2 q_i}{\partial x_k \partial x_k} + \frac{1}{3} \frac{\partial^2 q_k}{\partial x_i \partial x_k} \right) = -\frac{1}{\tau_R} q_i. \tag{11}$$

## 2.3 Conventional iterative scheme

In this section, the CIS is introduced to solve the phonon BTE for the steady-state solution. When the distribution function is known at the $n$-th iteration step, its value at the next iteration step can be obtained from the following equation,

$$\boldsymbol{v} \cdot \frac{\partial e^{n+1}}{\partial \boldsymbol{x}} = \frac{e_R^n - e^{n+1}}{\tau_R} + \frac{e_N^n - e^{n+1}}{\tau_N}, \tag{12}$$

where the spatial derivative can be approximated by the second-order upwind scheme, (the derivative in the layer near boundary approximated by first-order upwind scheme). The equilibrium distribution function $e_R^n$ and $e_N^n$ are calculated by the corresponding macroscopic quantities in the $n$-th iteration step:

$$e_R^n = e_{eq}^R(T^n), \quad e_N^n = e_{eq}^N(T^n, \boldsymbol{q}^n, \boldsymbol{v}). \tag{13}$$

The iteration repeats until the relative difference in temperature between two consecutive iterations are less than a fixed value.

We adopt the Fourier stability analysis to investigate the efficiency of CIS, that is, to see how fast the error decays when $n$ increases. We define the error functions between distribution functions at two consecutive iterations as[22]:

$$Y^{n+1}(\boldsymbol{x}, \boldsymbol{s}) = e^{n+1}(\boldsymbol{x}, \boldsymbol{s}) - e^n(\boldsymbol{x}, \boldsymbol{s}), \tag{14}$$

and the error functions for macroscopic quantities $M = [T, \boldsymbol{q}]$ between two consecutive iteration steps as:

$$\begin{aligned}\Phi_M^{n+1}(\boldsymbol{x}) &\equiv \left[\Phi_T^{n+1}, \Phi_q^{n+1}\right] = M^{n+1}(\boldsymbol{x}) - M^n(\boldsymbol{x}) \\ &= \int Y^{n+1}(\boldsymbol{x}, \boldsymbol{s}) \phi(\boldsymbol{v}) d\Omega,\end{aligned} \tag{15}$$

where

$$\phi(\boldsymbol{v}) = \left[\frac{1}{C_v}, v_1, v_2, v_3\right], \tag{16}$$

From Eq.(12), it can be easily found that $Y^{n+1}(\boldsymbol{x}, \boldsymbol{s})$ satisfies,

$$\boldsymbol{v} \cdot \nabla Y^{n+1} + \frac{1}{\tau_C} Y^{n+1} = \frac{C_v}{4\pi\tau_C} \Phi_T^n + \frac{3}{4\pi\tau_N} \boldsymbol{v} \cdot \Phi_q^n, \tag{17}$$

where $\tau_C$ is defined as

$$\frac{1}{\tau_C} = \frac{1}{\tau_R} + \frac{1}{\tau_N}. \tag{18}$$

To determine the error decay rate $\omega$, we perform the Fourier stability analysis by seeking the eigenfunctions $y(\boldsymbol{v})$ and $\alpha_M = [\alpha_T, \alpha_q]$ of the following forms:

$$Y^{n+1}(\boldsymbol{x}, \boldsymbol{s}) = \omega^n y(\boldsymbol{v}) \exp(i\boldsymbol{\theta} \cdot \boldsymbol{x}), \quad \Phi_M^{n+1}(\boldsymbol{x}) = \omega^{n+1} \alpha_M \exp(i\boldsymbol{\theta} \cdot \boldsymbol{x}), \tag{19}$$

where $\boldsymbol{\theta} = (\theta_1, \theta_2, \theta_3)$ is the wavevector of perturbance and $i$ is the imaginary unit. The iteration is unstable when the error decay rate is larger than unity, while slow (fast) convergence occurs when the error decay rate $|\omega|$ approaches one (zero). Note that the two exponents in the right-hand-side of Eq.(19) are different, due to the fact that in CIS we first need macroscopic quantities to start the iteration.

The combination of Eqs.(17) and (19) leads to,

$$y(\boldsymbol{v}) = \frac{\frac{C_v}{4\pi}\alpha_T + \frac{3\tau_C}{4\pi\tau_N}\boldsymbol{v}\cdot\boldsymbol{\alpha}_q}{1 + i\tau_C\boldsymbol{v}\cdot\boldsymbol{\theta}}. \tag{20}$$

In the following paper we assume the wave vector of perturbation satisfies $|\boldsymbol{\theta}| = 1$. Although in reality the perturbation may have various values of $|\boldsymbol{\theta}|$, the error decay rate does not change when the value of $\tau_C|\boldsymbol{\theta}|$ is fixed.

According to the definition of temperature and heat flux, we have

$$\begin{aligned}\Phi_T^n &= T^n - T^{n-1} = \frac{1}{C_v}\int Y^n d\Omega,\\ \Phi_q^n &= \boldsymbol{q}^n - \boldsymbol{q}^{n-1} = \int \boldsymbol{v} Y^n d\Omega,\end{aligned} \tag{21}$$

Because the problem is isotropic, we set $\theta_3 = 0$, and hence the problem is effectively two-dimensional. Combining Eqs.(19), (20), and (21), we obtain three linear algebraic equations for three unknown elements in $\alpha_M$. These algebraic equations can be written in the matrix form as $\omega\alpha_M^T = C\alpha_M^T$, where the superscript $T$ is the transpose operator, and $3\times 3$ matrix is:

$$C = \begin{pmatrix} c_1\int s_1 d\Omega & c_2\int v_1 s_1 d\Omega & c_2\int v_2 s_1 d\Omega \\ c_1\int v_1 s_1 d\Omega & c_2\int v_1^2 s_1 d\Omega & c_2\int v_1 v_2 s_1 d\Omega \\ c_1\int v_2 s_1 d\Omega & c_2\int v_1 v_2 s_1 d\Omega & c_2\int v_2^2 s_1 d\Omega \end{pmatrix}, \tag{22}$$

where

$$c_1 = \frac{C_v}{4\pi}, \quad c_2 = \frac{3\tau_C}{4\pi\tau_N}, \quad s_1 = \frac{1}{1 + i\tau_C\boldsymbol{v}\cdot\boldsymbol{\theta}}. \tag{23}$$

The error decay rate $\omega$ can be obtained by numerically computing the eigenvalues of the matrix $C$ and taking the maximum absolute value of $\omega$; results as a function of $\tau_R$ with different values of $\tau_N$ are shown in Figure 1. Note that in the numerical calculation we have assumed $|\boldsymbol{v}| = 1$. It is clear that when both $\tau_R$ and $\tau_N$ are large, $\omega$ goes to zero so that the error decays quickly. This means that CIS is quite efficient in ballistic regime. However, when either $\tau_R$ or $\tau_N$ is small, the convergence rate $\omega \to 1$, which indicates that the CIS works inefficiently in the diffusive/hydrodynamic regime.

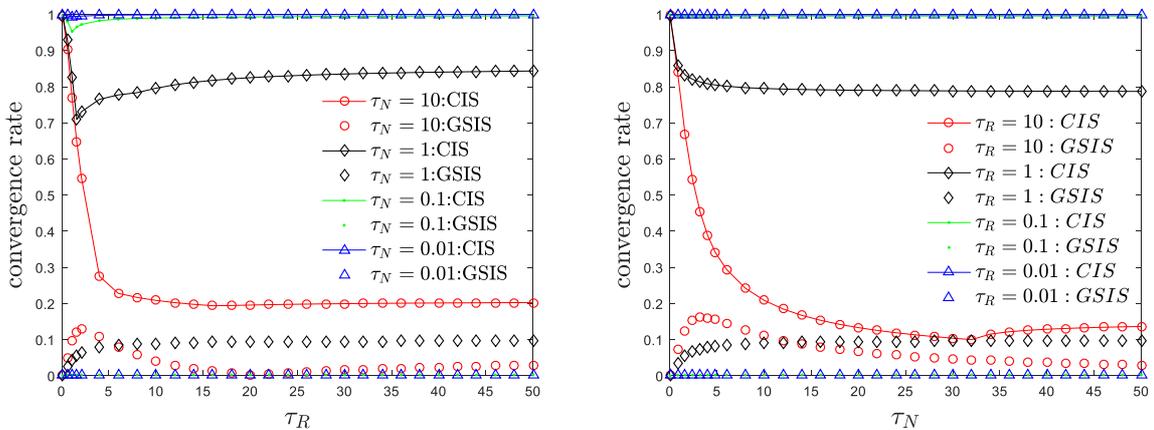

Figure 1 The convergence rate $\omega$ as a function of $\tau_R$ and $\tau_N$ in both CIS and GSIS.

## 2.4 General synthetic iterative scheme

To expedite the convergence of CIS, that is, to reduce the iteration numbers $n$ needed to obtain converged solutions, macroscopic synthetic equations are designed to guide the evolution of equilibrium distribution function for resistive and normal scatterings. That is, the phonon BTE and its moment equations are coupled and solved simultaneously at each iteration: the BTE provides high-order moments to moment equations, meanwhile, the moment equations provide macroscopic quantities appearing in the collision term.

First, in GSIS, given the value of distribution function $e^n$ at the $n$-th iteration step, its value at the intermediate $(n + 1/2)$-th step is obtained in a similar way to CIS:

$$\boldsymbol{v} \cdot \frac{\partial e^{n+\frac{1}{2}}}{\partial \boldsymbol{x}} = \frac{e_R^n - e^{n+\frac{1}{2}}}{\tau_R} + \frac{e_N^n - e^{n+\frac{1}{2}}}{\tau_N}. \tag{24}$$

And this distribution function $e^{n+1/2}$ will be used to close the high-order constitutive relations in the synthetic equations (10); when the synthetic equations are solved to obtain macroscopic quantities, say, $M^{n+1} = [T^{n+1}, \boldsymbol{v}^{n+1}]$, they will be used for next iteration, until convergence criterion is met.

Certainly, the macroscopic synthetic equations should be derived exactly from the phonon BTE. Here, the moment method is adopted. For simplicity, we consider the one-dimensional heat conduction in a slab as an example, see the geometry in Figure 2. In this case, Eq.(10) can be rewritten as the following form:

$$\frac{\partial q}{\partial x} = 0,$$
$$\frac{C_v}{3} \tau_R \frac{\partial T}{\partial x} - 2\pi \tau_C \tau_R \frac{\partial^2}{\partial x^2} \int_{-1}^{1} \left(v^2 - \frac{2}{5}\right) vedv = -q. \tag{25}$$

On integrating the second equation of Eq.(25) with respect to $x$ from 0 to 1, and combining the symmetry of this problem, we find the temperature, after the spatial discretization, can be expressed as:

$$\frac{C_v}{3} \tau_R T_i^{n+1} - 2\pi \tau_C \tau_R \frac{HOT_{i+1}^n - HOT_{i-1}^n}{x_{i+1} - x_{i-1}} = -q_*^n \left(x_i - \frac{1}{2}\right), \tag{26}$$

where

$$HOT(x) = \int_{-1}^{1} \left(v^2 - \frac{2}{5}\right) vedv, \quad \frac{\partial HOT(x)}{\partial x} = Q(x), \tag{27}$$

and

$$q_* = -\frac{C_v}{3} \tau_R (T_{N_x} - T_1) + 2\pi \tau_C \tau_R (Q_{N_x} - Q_1). \tag{28}$$

Meanwhile, the heat flux is updated as $q^{n+1} = q_*^n$.

To obtain the convergence rate of GSIS, we define the error functions:

$$Y^{n+\frac{1}{2}}(\boldsymbol{x}, \boldsymbol{s}) = e^{n+\frac{1}{2}}(\boldsymbol{x}, \boldsymbol{s}) - e^n(\boldsymbol{x}, \boldsymbol{s}) = \omega^n y(\boldsymbol{v}) \exp(i\boldsymbol{\theta} \cdot \boldsymbol{x}),$$
$$\Phi_M^{n+1}(\boldsymbol{x}) = M^{n+1}(\boldsymbol{x}) - M(\boldsymbol{x}) = \omega^{n+1} \alpha_M \exp(i\boldsymbol{\theta} \cdot \boldsymbol{x}), \tag{29}$$

Note that the difference between CIS and GSIS is that the macroscopic quantities $M^{n+1}$ are calculated from the synthetic equations, rather than directly from the distribution function $e^{n+1/2}$.

Through some simple algebraic operations, the error decay rate is found to satisfy the following $3 \times 3$ linear systems:

$$\omega \underbrace{\begin{pmatrix} 0 & i\theta_1 & i\theta_2 \\ \frac{C_v}{3}i\theta_1 & \frac{2\tau_C}{5}\theta_1^2 + \frac{\tau_C}{5}\theta_2^2 + \frac{1}{\tau_R} & \frac{\tau_C}{5}\theta_1\theta_2 \\ \frac{C_v}{3}i\theta_2 & \frac{\tau_C}{5}\theta_1\theta_2 & \frac{2\tau_C}{5}\theta_2^2 + \frac{\tau_C}{5}\theta_1^2 + \frac{1}{\tau_R} \end{pmatrix}}_{L} \begin{pmatrix} \alpha_T \\ \alpha_{q_1} \\ \alpha_{q_2} \end{pmatrix} =$$

$$\underbrace{\begin{pmatrix} 0 & 0 & 0 \\ c_1 \int s_1 y_1 d\Omega & c_2 \int v_1 s_1 y_1 d\Omega & c_2 \int v_2 s_1 y_1 d\Omega \\ c_1 \int s_2 y_1 d\Omega & c_2 \int v_1 s_2 y_1 d\Omega & c_2 \int v_2 s_2 y_1 d\Omega \end{pmatrix}}_{R} \begin{pmatrix} \alpha_T \\ \alpha_{q_1} \\ \alpha_{q_2} \end{pmatrix}, \quad (30)$$

where

$$c_1 = \frac{C_v}{4\pi}, \quad c_2 = \frac{3\tau_C}{4\pi\tau_N}, \quad y_1 = \frac{1}{1 + i\tau_C \boldsymbol{v} \cdot \boldsymbol{\theta}},$$

$$s_1 = -\tau_C \left[ \theta_1^2 \left(v_1^2 - \frac{2}{5}\right) v_1 + \theta_1\theta_2 v_1^2 v_2 + \theta_1\theta_2 \left(v_1^2 - \frac{1}{5}\right) v_2 + \theta_2^2 \left(v_2^2 - \frac{1}{5}\right) v_1 \right], \quad (31)$$

$$s_2 = -\tau_C \left[ \theta_1^2 \left(v_1^2 - \frac{1}{5}\right) v_2 + \theta_1\theta_2 \left(v_2^2 - \frac{1}{5}\right) v_1 + \theta_1\theta_2 v_1 v_2^2 + \theta_2^2 \left(v_2^2 - \frac{2}{5}\right) v_2 \right].$$

Equations (30) can be rewritten as the form of $L\omega\alpha_M^T = R\alpha_M^T$, and the error decay rate can be obtained from the eigenvalues of matrix $G = L^{-1}R$. A comparison of GSIS and CIS is shown in Figure 1, which clearly demonstrate that the GSIS great improves the iteration efficiency for small values of $\tau_R$ and $\tau_N$. However, for large relaxation times, the error decay rate goes to zero. Therefore, we modify the heat flux as

$$q^{n+1} = (1 - \beta)2\pi q^n + \beta q_*^n, \quad (32)$$

where the limiter is $\beta = \min(1, \tau_R)/\tau_R$. In this case, the error decay rate can be obtained by finding the eigenvalues of the matrix $\beta L^{-1}R + (1 - \beta)C$. Thus, fast convergence is achieved in the whole parameter region[22]. Numerical results in the following section demonstrates this choice not only allows fast convergence, but also enhances the stability of the new algorithm.

## 3. Numerical results and discussions

### 3.1 Numerical tests

In this section, the one-dimensional heat conduction across a dielectric film of thickness $L = 1$ is simulated, as shown in Figure 2. The temperatures on the boundaries located at $x = 0$ and $x = L$ are $T_L = -1/2$, $T_R = 1/2$, respectively. In the numerical simulation, the $x$ coordinate is normalized by the slab distance $L$. Therefore, the problem is characterized by the following Knudsen numbers:

$$Kn_R = \frac{|\boldsymbol{v}|\tau_R}{L}, \quad Kn_N = \frac{|\boldsymbol{v}|\tau_N}{L}. \quad (33)$$

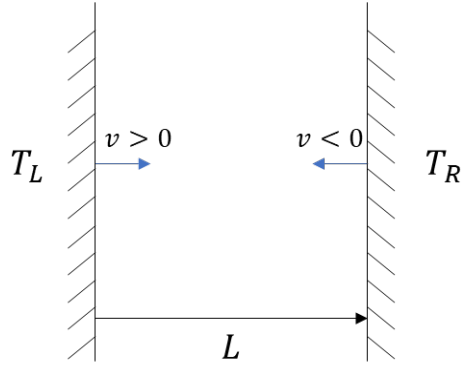

Figure 2 Schematic of one-dimensional heat conduction in a slab.

In order to solve the BTE, the thermalization boundary condition is adopted, in which a phonon is absorbed as it strikes the boundary, and a new phonon in equilibrium state with the boundary temperature is emitted into the computational domain. Therefore, the phonon reflected from the boundary can be expressed as,

$$e(\boldsymbol{x}_{BC}, \boldsymbol{s}) = e_{eq}^{R}(T_{BC}), \quad \boldsymbol{s} \cdot \boldsymbol{n} > 0, \tag{34}$$

where $BC$ is the boundary interface, $\boldsymbol{n}$ is the unit normal vector of the $BC$ pointing to the computational domain, and $T_{BC}$ is the temperature of $BC$.

We use 60 discrete Gauss-Legendre quadrature points in the $v_1 = |\boldsymbol{v}|\cos\theta$ direction so that the accuracy of the numerical quadrature with respect to $v_1$ can be ensured for all Knudsen numbers. The computational domain is discretized into 180 non-uniform cells,

$$x = d^3(10 - 15d + 6d^2), \tag{35}$$

where $d = (0,1,\ldots,N_x)/(N_x - 1)$ with $N_x = 181$, and $C_v$ is set to be 1. The solution is assumed to be converged when

$$\epsilon = \int_0^1 \frac{|T^{n+1} - T^n|}{T^n} dx < 1.0 \times 10^{-5}. \tag{36}$$

Figure 3 shows the numerical results predicted by the CIS and GSIS in different phonon transport regimes, i.e., different combinations of $Kn_R$ and $Kn_N$. Together with Table 1, it can be seen that while the GSIS yields exactly the same results as CIS in all flow regimes, it needs considerably less iteration steps in the diffusion/hydrodynamic regime.

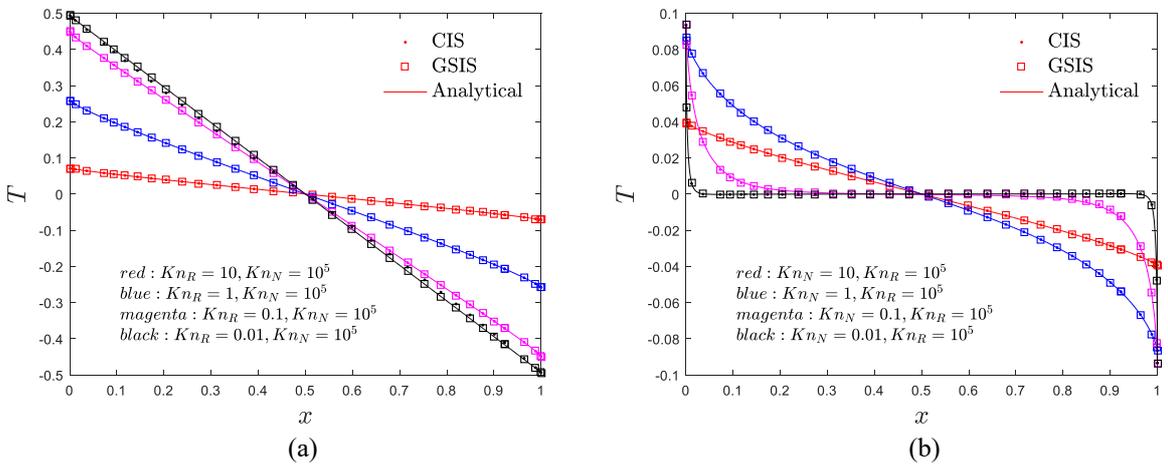

(a)         (b)

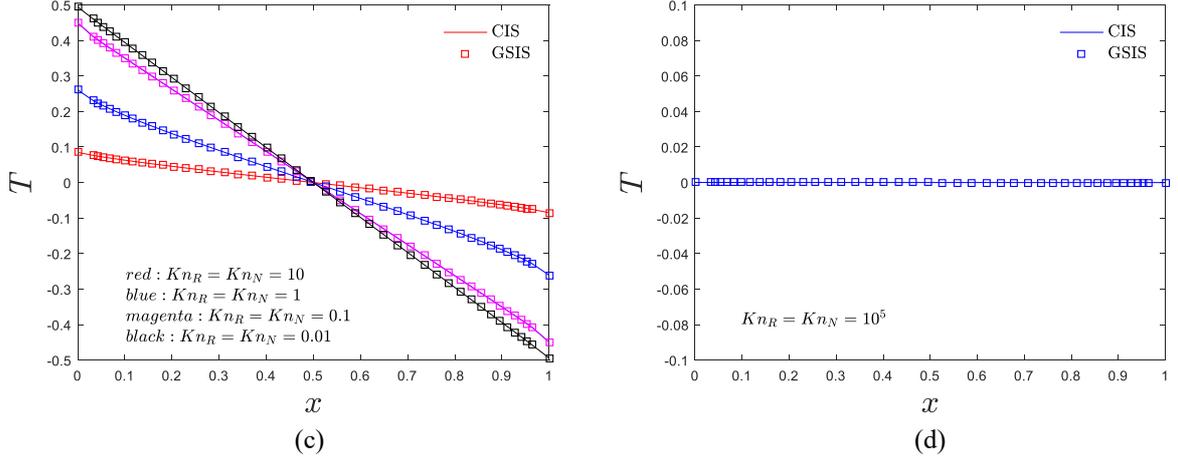

Figure 3 Temperature profiles solved by the CIS and GSIS in different regime. (a) Diffusive regime, (b) phonon hydrodynamic regime, (c) Ziman regime and (d) ballistic regime.

Table 1 The iteration number $n$ of CIS and GSIS under different combination of $\tau_R$ and $\tau_N$.

| $\tau_R$ | 10 | 1 | 0.1 | 0.01 | $10^5$ | $10^5$ | $10^5$ | $10^5$ | 10 | 1 | 0.1 | 0.01 |
|---|---|---|---|---|---|---|---|---|---|---|---|---|
| $\tau_N$ | $10^5$ | $10^5$ | $10^5$ | $10^5$ | 10 | 1 | 0.1 | 0.01 | 10 | 1 | 0.1 | 0.01 |
| $n_{CIS}$ | 4 | 9 | 83 | 25555 | 5 | 13 | 69 | 608 | 5 | 12 | 141 | 4025 |
| $n_{GSIS}$ | 5 | 8 | 17 | 19 | 6 | 11 | 14 | 15 | 6 | 8 | 17 | 19 |

### 3.2 Analysis of non-Fourier phenomenon

We now explore how the relaxation times of resistive and normal scatterings influence the heat conduction in materials described the Callaway model. The heat flux solved by GSIS can be written in the form of Fourier's heat conduct law, by introducing the effective thermal conductivity $\kappa_{eff}$,

$$q_{num} = -\kappa_{eff} \frac{\partial T}{\partial x}. \tag{37}$$

Meanwhile, we can derive the analytical thermal conductivity in the diffusive regime if only the resistive scattering is considered:

$$\kappa_{bulk} = \frac{1}{3} C_v \tau_R. \tag{38}$$

Then, the ratio of $\kappa_{bulk}$ and $\kappa_{eff}$ can reflect whether the Fourier's law is valid. It can easily know that only when $\kappa_{bulk}/\kappa_{eff} \to 1$, the Fourier's law can well predict the heat transfer.

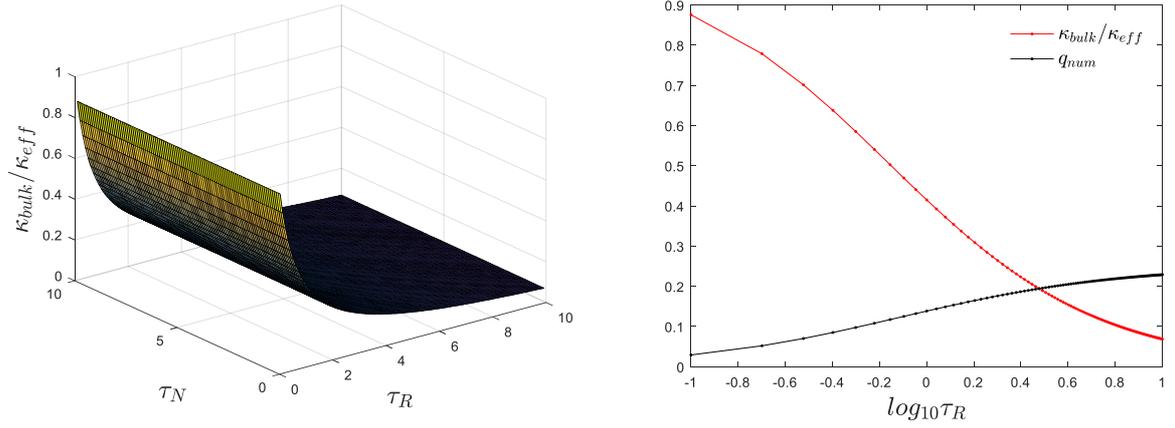

Figure 4 The ratio between the analytical and effective thermal conductivities as a function of $\tau_R$ and $\tau_N$.

Figure 4 shows that as $\tau_R$ decreases, the ratio of the analytical thermal conductivity to the efficient thermal conductivity $\kappa_{bulk}/\kappa_{eff}$ fall from 1 to 0. But when resistive relaxation time $\tau_R$ is fixed, the value of $\kappa_{bulk}/\kappa_{eff}$ almost remains unchanged when the normal relaxation time $\tau_N$ changes, which means that $\tau_N$ has little influence on the heat flux in this steady-state slab heat conduction. This is due to the fact that the normal scattering does not produce thermal resistance as the momentum is conserved.

However, the normal relaxation time affects the temperature field. In Figure 5(a), we see that when $\tau_N$ and $\tau_R$ is large, the phonon transport is in the ballistic regime where the heat conduction is mainly determined by phonon scattering at the slab boundaries, so the temperature appears to be the average of two wall temperatures, i.e., T=0. However, as $\tau_N$ decreases, the temperature deviates far from linear distribution that the temperature variation near boundary layer has a huge jump. The width of the strong temperature variation is about one mean free path $|v|\tau_N$ of the normal scattering, which is far smaller than the slab width. Therefore, when the explicit time-marching method is used, the CFL condition will requires a very small time step, and the numerical efficiency will be greatly reduced. From Figure 5(b), it can be known that as the resistive process get enhanced (i.e., when the value of $\tau_N$ is reduced), frequent energy exchange between phonons makes the temperature jump near the boundary reduced. So, the temperature returns to linear distribution.

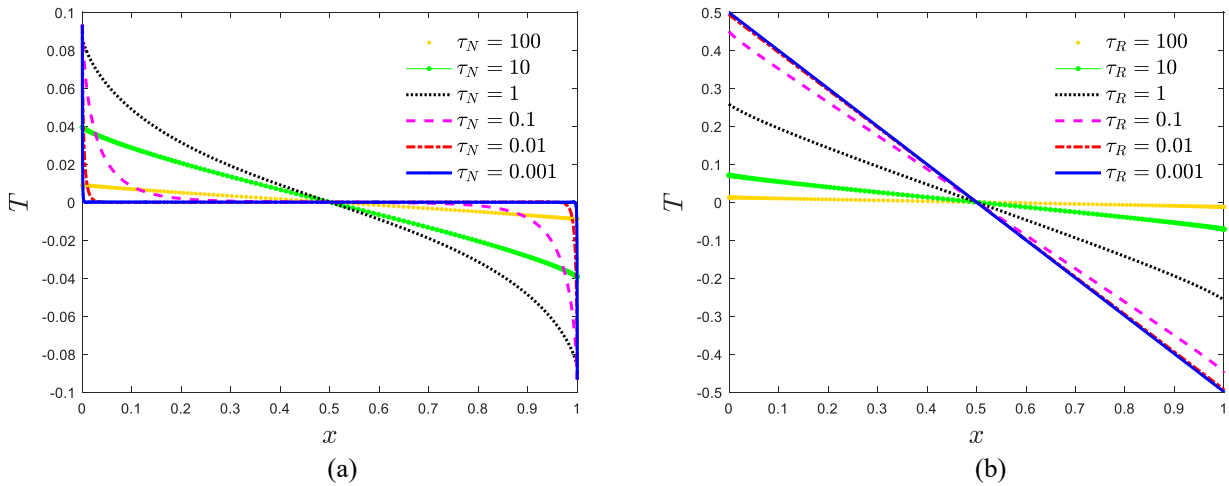

Figure 5 (a)Temperature across a film with different $\tau_N$, when $\tau_R = 10^5$; (b) Temperature across a film with different $\tau_R$, when $\tau_N = 10^5$.

## 4. Conclusion

We have developed a general synthetic iterative scheme to solve the steady-state problem of phonon Boltzmann equation efficiently and accurately. It can be found that the conventional iteration scheme works inefficiently when the Knudsen number is small, so we have introduced macroscopic equations to significantly improve its convergence to steady-state solutions. The proposed scheme solves mesoscopic Boltzmann equation and macroscopic moment equations simultaneously at each iteration. For the reason that the BTE provides high-order moment closure for moment equations, meanwhile, the macroscopic moment equations can also accelerate flow field signal exchange and modify distribution function, fast convergence in the entire range of Knudsen numbers can be achieved.

Several numerical tests in different phonon transport regimes have demonstrated that the GSIS predicts excellent results for all Knudsen numbers, and in the diffusion/hydrodynamic regime the number of iterations is reduced by several orders of magnitude. Furthermore, the Fourier stability analysis has been applied to find the convergence rate for both CIS and GSIS, which rigorously proves the efficiency of GSIS over CIS. In future works, we plan to explore the GSIS under two- and three-dimension. Additionally, as the gray matter assumption is adopted here, the relaxation times related to frequency will be considered in the future.